# QUANTUM-MECHANICAL CONSEQUENCES OF FIVE-DIMENSIONAL RELATIVITY

Paul S. Wesson

1. Department of Physics and Astronomy, University of Waterloo, Waterloo, Ontario N2L 3G1, Canada

2. Herzberg Institute of Astrophysics, National Research Council, Victoria, B.C. V9E 2E7, Canada.

Abstract: I outline a model where a massive particle in 4D spacetime follows a null or photon-like path in 5D canonical space. This leads to wave-particle duality, quantization and a Heisenberg-like uncertainty relation, along with other effects which show that it is possible to unify general relativity and quantum mechanics in 5D.

Correspondence: mail to (1) above; email = psw.papers@yahoo.ca

# QUANTUM-MECHANICAL CONSEQUENCES OF FIVE-DIMENSIONAL RELATIVITY

1. <u>Introduction</u>

In the search for a unified theory of gravity and particle physics via higher dimensions, a common assumption is that the base metric is an *N*-dimensional version of Minkowski space $(M_N)$. However, there is no evidence that this is the case. Indeed, in 4D general relativity most realistic solutions have metrics where the 3D part corresponding to ordinary space has spherical (not planar) symmetry. And in particle physics, most progress has been made by applying deSitter space, which is 3D spherically-symmetric and describes a vacuum with a cosmological constant $\Lambda$. These comments bear on the nature of the base metric for Kaluza-Klein theory, which in its modern non-compactified form is the simplest extension of Einstein theory. In 5D, there are good reasons to believe that the base metric is not $M_5$ but $C_5^*$ the so-called pure-canonical metric. In this, 4D spacetime is regarded as a 'spherically-symmetric' subspace in a 5D manifold whose extra part is flat. Important technical results have become available on this metric in recent years [1 – 4]. Principally, *null* geodesics in this 5D space correspond to the timelike geodesics of massive particles in 4D spacetime. And it is a corollary of Campbell's embedding theorem that $C_5^*$ smoothly embeds *all* solutions of Einstein's field equations of general relativity which are empty of ordinary matter but have a finite cosmological constant. In what follows, I wish to outline the main implications of 5D canonical space for particle dynamics.

The consequences of $C_5^*$ are surprisingly diverse. After establishing the properties of this metric in Section 2, I will treat wave-particle duality, quantization and uncertainty in Sections 3, 4 and 5. (Workers conversant with the 5D canonical metric may prefer to just skim Section 2.) Then in the conclusion of Section 6 I will suggest that these results cannot be fortuitous, but rather indicate that the 5D canonical metric describes the real base space of the world.

The nomenclature is standard. Letters A, B run 0, 123,4 for time, ordinary space and the extra dimension. The latter is denoted $x^4 = l$ to avoid confusion with other usages. Letters $\alpha, \beta$ run 0, 123 for spacetime. The constants $G$, $c$, $h$ are absorbed, unless they are needed to clarify a physical result.

2. <u>Canonical Space</u>

A general 5D line element $dS^2 = g_{AB} dx^A dx^B$ has 15 independent components of the metric tensor, which can in principle depend on the 4 coordinates of spacetime and the fifth one: $g_{AB} = g_{AB}(x^\gamma, l)$. Of these, 10 are identified with the gravitational potentials of general relativity in the spacetime limit, $g_{\alpha\beta}(x^\gamma)$. Of the others, the 4 off-diagonal ones are commonly identified with the potentials of electromagnetism in the same limit, $g_{4\alpha} \equiv A_\alpha(x^\gamma)$. The remaining coefficient is sometimes identified with the scalar field believed to be responsible for the rest masses of particles, $g_{44} = \pm\Phi^2(x^\gamma, l)$. Here both signs are admissible, because the extra coordinate does not have the physical nature of a time, so there is no problem with closed timelike paths when the signature is $(+---+)$, and numerous exact solutions of the 5D field equations are known with both signs. However, in 5D there are 5 degrees of coordinate freedom which are available to simplify the metric. These are often used to set $g_{4\alpha} \to 0$ and $|g_{44}| \to 1$. The resulting metric is still general as long as $g_{\alpha\beta} = g_{\alpha\beta}(x^\gamma, l)$; though the price paid for algebraic simplification of the metric is that the 4D sector may now harbor physical effects of electromagnetic and scalar type. Further simplification of the metric may be achieved by rewriting $g_{\alpha\beta}(x^\gamma, l)$ as $(l/L)^2 g_{\alpha\beta}(x^\gamma, l)$. Here $L$ is a constant length introduced for the consistency of physical dimensions (units), assuming $x^4 = l$ is taken to be a length coordinate. In general, $L$ measures the size of the potential well for the 4D part of the metric. When there is no ordinary matter present, it is known that $L$ measures the effects of the vacuum, which in general relativity are related to the cosmological constant $\Lambda$ (see below). The vacuum case corresponds to the special form of the 5D metric where the 4D part has the form $(l/L)^2 g_{\alpha\beta}(x^\gamma \text{ only})$. To distinguish this special case from the general one, it is referred to as the *pure*-canonical metric. It is essentially a 5D metric where the concept of spherical symmetry is extended from 3D space to 4D spacetime, and the latter describes any Einstein space with finite $\Lambda$. The simplest example is provided by the local version of deSitter space with $\Lambda > 0$ (to match the dark energy of cosmology), or anti-deSitter space with $\Lambda < 0$ (to match the vacuum fields of particles). The embedding of both kinds of 4D deSitter space in 5D has been well studied [5]. For present purposes, the pure-canonical or $C_5^*$ metric may be written

$$dS^2 = (l/L)^2 ds^2 \pm dl^2 \qquad (1.1)$$

$$ds^2 = g_{\alpha\beta}(x^\gamma) dx^\alpha dx^\beta \qquad (1.2)$$

$$\Lambda = \mp 3/L^2 \qquad (1.3)$$

Here a timelike extra coordinate corresponds to $\Lambda < 0$ while a spacelike one corresponds to $\Lambda > 0$.

There is an extensive literature on (1), which is reviewed in [4]. There is also a significant literature on the so-called *shifted* $C_5^*$ metric. This arises because the last term in (1.1) is unchanged by the shift $l \to (l - l_0)$ where $l_0$ is a constant. This amounts to shifting the locus of the 4D 'circle' on which spacetime is located, in a way analogous to how the 2D circle is changed from $r^2 d\theta^2$ to $(r + r_0)^2 d\theta^2$, where $\theta$ is an angular coordinate. [The analog of $r^2 d\theta^2$ in (1.1) is $l^2 d\theta^2$ where $d\theta \equiv ds/L$, where the 4D Ricci or scalar curvature is $R = -4\Lambda$, so for $\Lambda < 0$ the surface of spacetime $s$ is a hypersphere which is circumnavigated as $\theta$ runs from 0 to $2\pi$.] The importance of shifted $C_5^*$ is that the cosmological 'constant' becomes a function of the extra coordinate $x^4 = l$ [3]. The metric may be written

$$dS^2 = \left(\frac{l - l_0}{L}\right)^2 ds^2 \pm dl^2 \qquad (2.1)$$

$$ds^2 = g_{\alpha\beta}(x^\gamma) dx^\alpha dx^\beta \qquad (2.2)$$

$$\Lambda = \mp \frac{3}{L^2} \left( \frac{l}{l - l_0} \right)^2 \quad . \tag{2.3}$$

The change in the expression for $\Lambda$, here versus (1) above, may seem odd. However, it should be recalled that $\Lambda$ is a 4D quantity, evaluated as (1.3) or (2.3) by reducing the 5D field equations to their 4D counterparts. But any 4D quantity will in general change form under a coordinate change that involves $x^4 = l$. This because the 5D group of coordinate changes $x^A \to x^A(x^B)$ is broader than the 4D group $x^\alpha \to x^\alpha(x^\beta)$. In other words, $\Lambda$ is gauge dependent in 5D. This property may be used to model the big bang singularity and the resonances of elementary particles [6]. In general, the *l*-dependence of $\Lambda$ can obviously provide a route to the resolution of the cosmological-'constant' problem [7], since in 5D $|\Lambda|$ may be small on the macroscopic scale and large on the microscopic scale.

The focus of the present work, however, is dynamics. And in this regard, I will use the metrics (1) and (2) with the condition that causality in 5D is specified by $dS^2 = 0$. It has been known for some time that null-paths in 5D correspond to timelike (or null) paths in 4D [8]. That is, $dS^2 = 0$ corresponds to $ds^2 \geq 0$. In this way, conventional causality in spacetime is maintained. It should be noted, though, that with this condition it is impossible to use the 5D proper time as a dynamical parameter; and that the logical best choice for massive particles is the *4D* proper time. This may appear to be straightforward. But then the analogs of straight paths in 5D are generally *curved* paths in 4D. Accelerations are therefore to be expected in 4D which are anomalous as judged from the viewpoint of conventional dynamics, and it is these which ultimately underlie the effects to be discussed below.

3. <u>Wave-Particle Duality</u>

There has never been a convincing explanation on the basis of 4D theory for why things like electrons can behave as particles *and* waves.

In 5D, however, an explanation is immediately apparent by applying the null-path condition $\left(dS^2 = 0\right)$ to the shifted pure-canonical metric (2.1). This gives

$$l = l_0 + l_* \exp(\pm s/L) \qquad (l \text{ spacelike}) \qquad (3.1)$$

$$l = l_0 + l_* \exp(\pm is/L) \qquad (l \text{ timelike}) \ . \qquad (3.2)$$

Here the sign choice merely reflects the reversibility of the motion in the extra dimension. The path $l(s)$ is defined by the shift constant $l_0$, another constant length $l_*$ and the size of the potential well $L$. For (3.1) the motion is monotonic, and the hypersurface which defines spacetime has curvature scalar $R = -4\Lambda < 0$ and is open, where in 3D there is a repulsive force because $\Lambda > 0$. For (3.2) the motion is oscillatory, spacetime has $R = -4\Lambda > 0$ and is closed, where in 3D there is an attractive force because $\Lambda < 0$. Obviously, (3.1) is particle-like while (3.2) is wave-like, the distinction depending on whether the motion is free or confined.

The role of $\Lambda$ here is critical. By (2.3), it is seen that $|\Lambda| \to 3/L^2$ for $l \to \infty$, so asymptotically $\Lambda$ has the constant value typical of the unshifted $C_5^*$ metric (1). But the shift causes $|\Lambda| \to \infty$ for $l \to l_0$, a result which in practice creates a kind of 'groove' in which the wave (3.2) is trapped. The amplitude of the wave is $l_*$ and its wavelength is $L$. The behaviour of $\Lambda$ is illustrated in Figure 1.

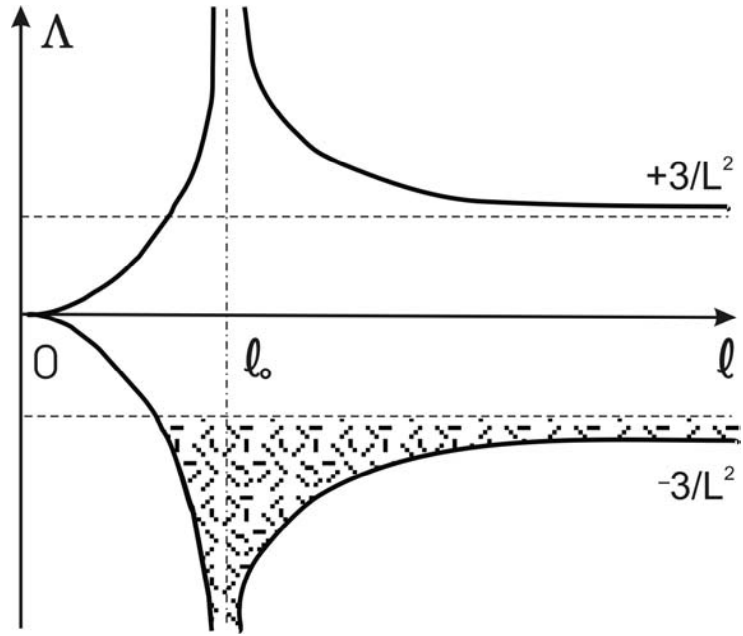

Figure 1. A schematic plot of the cosmological 'constant' $\Lambda$ as a function of the extra co-ordinate $l$, according to (2.3) of the text. The stippled region is a 'trap' for the wave (3.2).

It is instructive to compare the above with the picture for the unshifted metric (1). There are still two modes for $l(s)$, obtained by dropping $l_0$ from (3). But now there is no $\Lambda$-'groove' in which the wave of the oscillatory mode can be trapped. For this and other reasons, I think the shifted metric (2) is superior.

It is instructive to compare the dynamics for 5D canonical-type metrics like (1) with the properties of other metrics including $M_5$ [3,4]. By minimizing the 5D interval $S$ around zero and using the 4D proper interval *s* as parameter, the general equations of motion fall out as two sets. One of these is a set of 4 relations for spacetime, which comprises the usual geodesic motion found in Einstein's theory with an extra term that is due to the fifth dimension. The latter is a force per unit mass, or acceleration, which acts parallel to the 4-velocity $u^\alpha \equiv dx^\alpha/ds$ and is proportional to the relative velocity between the 4D and 5D frames measured by $dl/ds$. Its precise form is given most quickly by perturbing the standard normalization condition for the 4-velocities, now in the form $g_{\alpha\beta}(x^\gamma, l) u^\alpha u^\beta = 1$. This gives the extra force as

$$f^\mu = -\frac{u^\mu}{2}\left(u^\alpha u^\beta \frac{\partial g_{\alpha\beta}}{\partial l}\right)\frac{dl}{ds} \quad . \tag{4}$$

Here the term in parentheses represents the coupling between the 4D and 5D frames, and its value depends on how the 4D metric tensor involves $x^4 = l$. However, while $f^\mu$ of (4) will in general be present for an *l*-dependent spacetime, it may be shown that for the pure-canonical form it can be rendered zero [1, 10]. This is because the coupling between the 4D and 5D frames can be arranged to be zero by an appropriate choice of coordinates. Thus $C_5^*$ is special in that the 4D equations of motion are the *same* as in Einstein's theory. As for the motion in the extra dimension, it is given in general by the extra component of the 5D geodesic equation, though above it was derived directly from the metric by applying the null-path condition and is given by (3). It should be noted that the condition $dS^2 = 0$ obliges the choice of the 4D interval *s* as dynamical parameter even for 5D Minkowski space, with the result that the motion appears to be accelerated be-cause straight paths in *S* are curved paths in *s*. In fact the acceleration involved is given by (4). In $M_5$, the evolution of the 5 coordinates for spacelike *l* is given by $x^A = x_0^A \exp(\pm s/L)$, where $x_0^A$ are constants and the origin has been suitably defined. This motion is of the same type as (3.1) above, and will be taken up again below.

For now, we note that in the relation (3.2) the function for the wave behaviour of the extra coordinate *l* is similar to the 4D wave function $\psi$ in old wave mechanics. In fact, modulo the shift, the two motions are identical given the identification $L = h/mc$, the Compton wavelength of the particle with mass *m*. We thus infer that $\psi \sim l$, where in essence *l* is the coordinate conjugate to the mass *m*. This inference can be confirmed with some algebra: combining the components of the geodesic equation as discussed above, subject to the constraint that the 5D interval be null, I find a scalar equation in *l* which is identical to the Klein-Gordon equation in $\psi$, namely

$$\Box^2 \psi + m^2 \psi = 0 \quad . \tag{5}$$

Here $\Box^2 \psi \equiv g^{\alpha\beta} \psi_{,\alpha;\beta}$ where the comma and the semicolon denote the partial and covariant derivative (see ref. 4 for the details of the analysis). In a way, this correspondence might have been guessed from the form of (3); but the noted algebra is general, making the identification $l \sim \psi$ secure.

The preceding considerations make the correspondence between classical field theory and quantum mechanics quite clear: The 5D pure-canonical metric (2), which can embed any vacuum solution of Einstein's 4D equations with a cosmological 'constant', yields on applying the null-path condition the motions (3.1) and (3.2) which are particle-like and wave-like; and while the 4D the motion is in general affected by the extra force (4), this is zero for $C_5^*$ so that the evolution of timelike $l$ is equivalent to the Klein-Gordon equation (5) for the wave function. This correspondence means that wave-particle duality can be understood as the consequence of 5D Kaluza-Klein theory, but where the base metric is not the often-assumed Minkowski one with rectangular form, but rather the canonical one with a kind of super-spherically-symmetric form.

4. Quantization

Given that a canonical-type metric with signature $(+---+)$ describes an $l$-wave with wavelength $L = h/mc$, it is straightforward to show quantization.

This depends, however, on topology. For example, if the null 5D line element were assumed (erroneously) to be $M_5$, the relation $dS^2 = 0 = ds^2 \pm dl^2$ only leads to quantization of $ds$ if there is some kind of planar structure in the extra dimension such that $dl$ changes by discrete steps, a situation which is unnatural. The metric and the exact solution of the field equations which it represents do not completely inform about the topology. But as noted above, in the case of (2) and (3.2), the 4D curvature $R = -4\Lambda$ is positive, implying that the hypersurface $s$ of spacetime is closed. The simplest topology, and the appropriate one for anti-deSitter, is therefore circular [5]. This is illustrated in Figure 2.

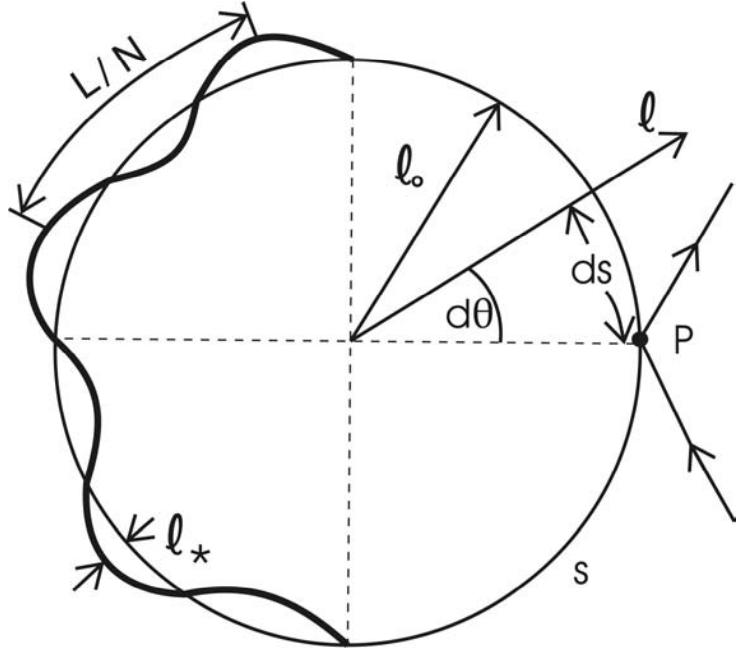

Figure 2. A schematic representation of the wave in the extra coordinate *l* as it propagates around the closed surface of spacetime *s*, according to (3.2) of the text. In the latter the fundamental mode is treated, whereas in the figure a harmonic with shorter wavelength is shown. The point *P* marks the impact of a scattered particle.

There, the locus of the *l*-wave is $l_0$ and the fundamental mode has wavelength $L$ (though overtones are presumably allowed with wavelengths $L/N$ where $N$ is an integer). Clearly, one wave has to fit into the circumference $2\pi l_0$, so $2\pi l_0 = L = h/mc$ and $l_0 = \hbar/mc$. This length fixes the angle $d\theta = ds/l_0$ around the orbit from a point $P$ which marks the impact of a scattered particle. Substituting the previously-found expression for $l_0$ gives the angle as $d\theta = mc\,ds/\hbar$. For $n$ revolutions,

$$\int_0^{2\pi n} d\theta = \int \frac{mcds}{\hbar},$$

$$mcs = nh \quad . \tag{6}$$

The analysis leading to this result is transparently simple.

Another simple result concerns the cosmological 'constant', which by (2.3) for $l \to \infty$ is $|\Lambda_\infty| = 3/L^2$. However, from above, $L = h/mc$. Therefore

$$|\Lambda_\infty| = 3\left(\frac{mc}{h}\right)^2 \quad . \tag{7}$$

This is to be interpreted as meaning that each particle has its 'own' value of this parameter, dependent on its mass.

5. <u>Uncertainty</u>

The previous two sections dealt with the wave-like behaviour of the extra coordinate in the canonical metric with signature $(+---+)$ and a 4D space that is closed. However, the opposite case involves a monotonic evolution as in (3.1) that affects the dynamics of free particles.

For $C_5^*$, the 4D line element $ds^2$ is modulated by a factor $l^2$, which evolves as $\exp(\pm 2s/L)$. So while the coordinates 'internal' to spacetime may not show it, there is in effect a kind of inflation or deflation of the 4D part of the 5D metric which is analogous to the deSitter expansion of the 3-space in inflationary cosmology [4]. The same kind of effect occurs for $M_5$ (see above), though there it is due to using the 4D $s$ rather than the 5D $S$ to parametize the dynamics [1,10]. In both cases, the effect may be quantified in terms of

$$X^\alpha = X_0^\alpha e^{\pm s/L} \qquad . \tag{8}$$

Here $X_0^\alpha$ are constants, the origin has been appropriately defined, and $L = h/mc$ from before. The question arises of what consequences follow from the fact that the motion is effectively accelerated.

In the present model, the mass of the test particle $m$ is constant, so the 4-momenta $P^\alpha \equiv mU^\alpha$ vary only because of the change in the 4-velocities $U^\alpha \equiv dX^\alpha/ds$. Then (8) gives

$$dX_\alpha = \pm(1/L) X_\alpha ds \tag{9.1}$$

$$dP^\alpha = (1/L^2) m X^\alpha ds \qquad . \tag{9.2}$$

The product of these, together with $X_\alpha X^\alpha = L^2$ and $L = h/mc$, yields

$$dP^\alpha dX_\alpha = \pm \left(\frac{mcds}{h}\right)^2 \frac{h}{c} = \pm \frac{h}{c} \qquad . \tag{10}$$

Here the usual quantization rule, derived in Section 4, has been used. It should be noted that $P^\alpha$ in (10) is defined with a dimensionless 4-velocity (which is the ordinary velocity divided by $c$); and that the 5D embedding of deSitter space is homogeneous and isotropic [5], so that (10) applies equally to each direction in spacetime. Also, the uncertainty specified by (10) is that due to the evolution of the $C_5^*$ embedding factor multiplied onto spacetime, and any other perturbations in the latter will add to the effect. That is, (10) represents the minimum level of uncertainty. This explains the much-discussed expressions $|\Delta E \Delta t| \geq h$ and $|\Delta p \Delta x| \geq h$ for the time and space directions of 4D relativity.

The Uncertainty Principle of Heisenberg has generated much philosophical discussion. However, while the effect is real, much of the discussion about its origin appears to be flummery. From the 5D perspective, uncertainty arises because 4D spacetime is affected by an evolving higher-dimensional factor, which in turn is due to the fact that the underlying geometry is canonical.

6. Conclusion

Many natural systems are described by 4D metrics whose 3D (ordinary space) parts are spherically symmetric. This geometry can be extended to 5D, where spacetime itself is a kind of spherically-symmetric subspace with the role of the radius played by the extra coordinate (Section 1). This canonical metric can be used to investigate null paths in 5D which correspond to the timelike paths of massive particles in 4D. Canonical space has been shown in Sections 3, 4 and 5 to yield straightforward accounts of wave-particle duality, quantization and uncertainty.

These subjects involve diverse physics, and it seems to me almost impossible that they flow from the $C_5$ metric by a series of algebraic accidents. Rather, it seems likely that $C_5$ is actually the 'base' metric of the world, as opposed to the $M_5$ often assumed. To the results derived above, concerning the dynamics of particles, should be added the successes of the canonical metric in application to cosmology [9]. There are also many exact solutions of the 5D field equations known which have metrics of $C_5$ type and await application to physics [10]. In regard to the simple model outlined above, attention should be given to the possibility of particle production in the $\Lambda$-well $(l \to l_0)$, the question of the entropy of the central 'object' and its area $(\pi l_0^2)$, the extension to 5D of the 4D wave function and probability distribution $(\psi\psi^* \sim ll^*)$, and the nature of harmonics of the fundamental wavelength $(L = h/mc)$. Notwithstanding the need to investigate these things, I believe that the particle model presented in this account is superior to others, and in particular to previous ones proposed by myself. In regard to more sophisticated models, attention should be given to extensions of the deSitter vacuum solution, the inclusion of matter, and the integration of the particle picture with the cosmological one.

As of now, it appears that a unified theory of gravity and the interactions of particles is achievable using Kaluza-Klein theory, provided that the base metric has canonical form.

Acknowledgements

This paper is a companion to the preceding one (ArXiv:1102.0801). I am grateful for discussions with B. Mashhoon and other members of the S.T.M. group (http://astro.uwaterloo.ca/~wesson). This work was partially supported by N.S.E.R.C.